\documentclass[12pt,a4paper]{article}
\usepackage{amssymb}

\def\beq{\begin{equation}}
\def\eeq{\end{equation}}

\def\be{\begin{displaymath}}
\def\ee{\end{displaymath}}

\newtheorem{thm}{Result}

\def\Tr{\mathop{\rm Tr}\nolimits}
\def\x {\stackrel {\textstyle \otimes}{,}}
\def\rank{\mathop{\rm rank}\nolimits}

\title{
A Conjectured R-Matrix}

\author{H. W. Braden\\
\normalsize
\em Department of Mathematics and Statistics,\\
\normalsize
\em The University of Edinburgh, \\
\normalsize
\em Edinburgh, UK \\
\normalsize
e-mail: hwb@ed.ac.uk
\\
}

\begin{document}

\renewcommand{\thepage}{}
\begin{titlepage}

\maketitle
\vskip-9.5cm
\hskip10.4cm
MS-97-013
\vskip.2cm
\hskip10.4cm
\sf solv-int/9710011 \rm
\vskip8.8cm

\begin{abstract}
\end{abstract}
A new spectral parameter independent R-matrix 
(that depends on all of the dynamical variables) 
is proposed for the elliptic Calogero-Moser models.
Necessary and sufficient conditions for this R-matrix to exist reduce
to an equality between determinants of matrices involving elliptic
functions. The needed identity appears new and is still unproven in
full generality: we present it as a conjecture.
\vfill
\end{titlepage}
\renewcommand{\thepage}{\arabic{page}}

\section{Introduction}
This paper is directed towards the construction of an R-matrix for the
elliptic Calogero-Moser models. Together with a Lax pair, the R-matrix is
a key ingredient of the modern approach to completely integrable 
systems.  In this approach the Lax equation $\dot L =\left[ L,M\right] $ 
enables us to
construct conserved quantities such as the traces $\Tr{L\sp k}$,
while the R-matrix shows these quantities Poisson commute. A system is said
to be completely integrable when we have enough independent, 
mutually Poisson commuting, conserved quantities and for such systems
R-matrices  must exist \cite{BV}. For completely integrable systems
Liouville's theorem \cite{Liouv,Arnold} tells us that we may integrate the
equations of motion by quadratures; with certain completeness 
assumptions\footnote{Flaschka \cite{flaschka} gives several simple
examples where these assumptions fail.}
on the flows, Arnold's extension of Liouville's theorem ensures the
existence of global action-angle variables. 
The R-matrix is  also an essential ingredient when examining the
separation of variables of such integrable systems \cite{Sk2, KNS}.

Recent work has yielded necessary and sufficient conditions for an R-matrix
to exist, together with an explicit construction, and we shall now apply
this to the elliptic Calogero-Moser models. For the rational and
trigonometric degenerations of these models Avan and Talon \cite{AT}
have constructed R-matrices under an assumption of momentum independence;
R-matrices can in principle be functions of the dynamical variables.
For the elliptic models however, the work of \cite{BS} shows that no
momentum independent R-matrices can be constructed for more than 3 particles.
This restriction can be circumvented by 
considering R-matrices depending on a spectral parameter and such
 momentum independent
R-matrices  were found for the elliptic Calogero-Moser models by
Sklyanin \cite{Sk1} and Braden and Suzuki \cite{BS}. A question however remains
unanswered: { are there spectral parameter independent R-matrices for the
elliptic Calogero-Moser models?}
Here we propose such R-matrices. The necessary and sufficient conditions for
the R-matrix to exist reduce to a single identity involving matrices
with elliptic function entries. This identity appears new and we have
been unable to prove it in generality: it is given here as a conjecture.

An outline of the paper is as follows. In sections two and three
we briefly review the construction of R-matrices and the Calogero-Moser models
respectively. In section four we combine this material to obtain necessary
and sufficient conditions for a spectral parameter independent R-matrix
for the elliptic Calogero-Moser models to exist, specifying the R-matrix
when such holds true. The necessary and sufficient conditions may be
expressed as an equality between two determinants involving
elliptic functions: this is presented in section five. The final
section is devoted to a brief discussion.

\section{The Construction}
The recent advances in the construction and understanding  of R-matrices 
are consequences of the study  \cite{Br1} of a more general matrix equation
\beq
A\sp{T}X-X\sp{T}A=B.
\label{matrixeqn}
\eeq
As we shall review, the R-matrix equation is a particular example of this.
Because $A$ is in general singular the general solution to (\ref{matrixeqn})
is in terms of a generalized inverse $G$ satisfying
\beq
AGA=A\quad\quad{\rm and}\quad\quad GAG=G.
\label{geninverse1}
\eeq
Such a generalized inverse\footnote{Accounts of generalized
inverses may be found in \cite{AG,Ca,Pr,RM}.
Indeed the  Moore-Penrose inverse
-which is unique and always exists- further
satisfies $(AG)\sp\dagger=AG$, $(GA)\sp\dagger=GA$.} always exists.
Given  a $G$ satisfying (\ref{geninverse1}) we  have at hand
projection operators $P_1=GA$ and $P_2=AG$ which satisfy
\beq
AP_1=P_2A=A,\quad\quad P_1G=GP_2=G.
\label{proj}
\eeq
The matrix equation (\ref{matrixeqn}) then has solutions if and only if
$$
\displaylines{
(C1)\quad\quad\phantom{xxxx} B\sp{T}=-B,\hfill \cr
(C2)\quad\quad (1-P_1\sp{T})B(1-P_1)=0,\hfill\cr}
$$
in which case the general solution is
\beq
X={1\over2} G\sp{T} B P_1+G\sp{T} B(1-P_1) +(1-P_2\sp{T})Y+ (P_2\sp{T}ZP_2)A
\label{gensoln}
\eeq
where $Y$ is arbitrary and $Z$ is only constrained by the requirement that
$P_2\sp{T}ZP_2$ be symmetric.
Although the general solution appears to depend on the  generalized
inverse $G$ any other choice of generalized inverse  will only change the 
solution within the ambiguities given by (\ref{gensoln}).

The classical R-matrix construction \cite{STS} arises as a particular case 
of (\ref{gensoln}) as follows.
Suppose the Lax matrix $L$ is in a representation $E$ of a Lie algebra 
${\mathfrak{g}}$ (here taken to be semi-simple). The classical 
R-matrix is a $E{\otimes }E$ valued matrix such that
\beq
[R, L\otimes 1]-[{R\sp{T}},1\otimes L] =\{L\x L\}.
\label{rmatrix}
\eeq
Let  $T_\mu$ denote a  basis for the  (finite dimensional)
Lie algebra ${\mathfrak{g}}$ 
with $[T_\mu,T_\nu]=c_{\mu \nu}\sp\lambda\ T_\lambda$ defining the 
structure constants of ${\mathfrak{g}}$. Set $\phi(T_\mu)=X_\mu$,  where
$\phi$ yields the
representation $E$ of the Lie algebra ${\mathfrak{g}}$; we may take this
to be a faithful representation. With 
$L= \sum_{\mu}L\sp\mu X_\mu $ the left-hand side of (\ref{rmatrix})
becomes
$$
\{L\x L\}=\sum_{\mu,\nu} \{ L\sp\mu, L\sp\nu \} X_\mu\otimes X_\nu
$$
while upon setting $R=R\sp{\mu\nu}X_\mu\otimes X_\nu$ and
${R\sp{T}}=R\sp{\nu\mu}X_\mu\otimes X_\nu$ the right-hand side
yields
\begin{eqnarray}
[R, L\otimes 1]-[{R\sp{T}},1\otimes L]&=&
R\sp{\mu\nu}([X_\mu,L]\otimes X_\nu-X_\nu\otimes[X_\mu,L])\cr
&=& R\sp{\mu\nu}L\sp\lambda
 ([X_\mu,X_\lambda]\otimes X_\nu-X_\nu\otimes[X_\mu,X_\lambda])\cr
&=&( R\sp{\tau\nu}c_{\tau\lambda}\sp\mu L\sp\lambda -
    R\sp{\tau\mu}c_{\tau\lambda}\sp\nu L\sp\lambda) X_\mu\otimes X_\nu.
\nonumber
\end{eqnarray}
By identifying 
$ A\sp{\mu\nu}= c_{\mu\lambda}\sp\nu L\sp\lambda \equiv-ad(L)\sp\nu _\mu$,
$ B\sp{\mu\nu}=\{ L\sp\mu, L\sp\nu \}$ and $ X\sp{\mu\nu}=R\sp{\mu\nu}$
we see that (\ref{rmatrix}) is an example of (\ref{matrixeqn}).

In the R-matrix context the matrix $B$ is manifestly antisymmetric because
of the antisymmetry of the Poisson bracket and so $(C1)$ is clearly
satisfied. We have thus reduced the existence of an R-matrix to the 
single consistency equation $(C2)$ and the construction of a generalized
inverse to $ad(L)$. 

The construction of a generalized inverse for (a generic) $ad(L)$  was given 
in \cite{Br2}.
Let $X_\mu$ denote a Cartan-Weyl basis for the Lie algebra ${\mathfrak{g}}$.
That is $\{X_\mu\}=\{H_i,E_\alpha\}$, where $\{H_i\}$ is a basis for the 
Cartan subalgebra ${\mathfrak{h}}$
and $\{E_\alpha\}$ is the set of  step operators (labelled by
the root system $\Phi$ of ${\mathfrak{g}}$). The structure constants
are found from 
$$
[H_i,E_\alpha ]=\alpha_i E_\alpha,\quad
[E_\alpha,E_{-\alpha}]= \alpha\sp\vee\cdot H\quad 
{\rm and}\quad
[E_\alpha,E_\beta ]=N_{\alpha,\beta}E_{\alpha+\beta}
\quad{\rm if}\ \alpha+\beta\in\Phi .
$$
Here $N_{\alpha,\beta}=c_{\alpha\, \beta }^{\alpha+ \beta }$.
With these definitions we then have that
\beq
\begin{array}{rrl}
& j\qquad\quad\beta\qquad\quad&\\
&\downarrow\qquad\quad\downarrow\qquad\quad&\\
\\
\left( adL\right) =& 
\begin{array}{c}
i\rightarrow \\ 
\alpha \rightarrow
\end{array}
\left( 
\begin{array}{cc}
0 & -\beta _{i}^{\vee }L^{-\beta } \\ 
-\alpha _{j}L^{\alpha } & \Lambda _{\beta }^{\alpha }
\end{array}
\right) &=\left( 
\begin{array}{cc}
0 & u^{T} \\ 
v & \Lambda
\end{array}
\right) 
\end{array}
\eeq
where we index the rows and columns first by the Cartan subalgebra basis
$\{i,j:1\ldots \rank\mathfrak{g}\}$ then the root system 
$\{\alpha,\beta \in \Phi\}$. We will use this block decomposition of
matrices throughout. Here $u$ and $v$ are 
$|\Phi|\times \rank\mathfrak{g}$ matrices and we have introduced the
$|\Phi|\times|\Phi|$ matrix
\beq
\Lambda _{\beta }^{\alpha }=
\alpha \cdot L\, \delta_{\beta }^{\alpha }+
c_{\alpha -\beta \beta }^{\alpha }\, L^{\alpha -\beta },
\label{Lambdadef}
\eeq
where $\alpha \cdot L=\sum_{i=1}\sp{\rank\mathfrak{g}}\alpha_i L\sp{i}$.
With these definitions we have \cite{Br2} that for
generic $L$ the matrix $\Lambda $ is invertible and a
 generalised inverse of $ad(L)$ is given by
\beq
\left( \begin{array}{cc}
1 & 0 \\ 
-\Lambda ^{-1}v & 1
\end{array}
\right) \left( 
\begin{array}{cc}
0 & 0 \\ 
0 & \Lambda ^{-1}
\end{array}
\right) \left( 
\begin{array}{cc}
1 & -u^{T}\Lambda ^{-1} \\ 
0 & 1
\end{array}
\right)
=\left( \begin{array}{cc}
0&0\\
0 & \Lambda ^{-1}
\end{array}
\right). 
\label{geninvadL}
\eeq

We may now assemble these results to yield the R-matrix. It is
convenient to express the Poisson brackets of the entries of $L$ in the same
block form in the Cartan-Weyl basis:
\beq
B=\left(
\begin{array}{cc}
\zeta & - \mu\sp{T} \\
\mu & \phi
\end{array}
\right)= -B ^{T}
\label{bdef}
\eeq
where $B\sp{\alpha j }=\{L\sp\alpha,L\sp{j}\}=\mu_{\alpha j }$ and so on.
From the fact that $A=-ad(L)\sp{T}$, a generalized inverse of $A$
is given by minus the transpose of the  generalized inverse (\ref{geninvadL})
and consequently we obtain the  projectors
$$
P_{1}=\left(
\begin{array}{cc}
0 & 0 \\
\Lambda ^{-1T}u & 1
\end{array}
\right)
,\quad\quad
P_{2}=\left(
\begin{array}{cc}
0 & v ^{T}\Lambda ^{-1T} \\
0 & 1
\end{array}
\right).
$$
The constraint $(C2)$ is now (the $\rank\mathfrak{g}\times\rank\mathfrak{g}$
matrix equation)
\beq
(C2)\quad
0= \left( 1-P_{1}^{T}\right) B\left(
1-P_{1}\right) \equiv
\zeta +\mu\sp{T} \Lambda ^{-1T} u - u\sp{T}\Lambda ^{-1} \mu +
u\sp{T}\Lambda ^{-1}\phi\, \Lambda ^{-1T} u.
\label{consgen}
\eeq

Supposing the constraint $(C2)$ is satisfied we then find
from (\ref{gensoln}) the general R-matrix takes the form
\beq
R=\left(
\begin{array}{cc}
0 & 0 \\
- \Lambda ^{-1}\mu +\frac{1}{2} \Lambda ^{-1}\phi\, \Lambda ^{-1T} u
& -\frac{1}{2} \Lambda ^{-1}\phi
\end{array}
\right) +\left(
\begin{array}{cc}
p & q \\
-\Lambda ^{-1}v p-Fu & -\Lambda ^{-1}v q-F\Lambda ^{T}
\end{array}
\right).
\label{rmat}
\eeq

The second term characterises the ambiguity in R where we have
parameterised the matrices $Y,Z$ in (\ref{gensoln}) by
$Y=\left(\begin{array}{cc}p&q\\ r&s \end{array} \right)$
and $Z=\left(\begin{array}{cc}a&b\\ c&d \end{array} \right)$.
Here the matrices $p,q$ are arbitrary 
while the entries of $Z$ are such that
\beq
F=\Lambda ^{-1}v a v ^{T}\Lambda ^{-1T}+d+\Lambda
^{-1}v b+c v ^{T}\Lambda ^{-1T}
\label{Fdef}
\eeq
is symmetric.

\section{The Calogero-Moser Models}

We now recall the salient features of the Calogero-Moser models and in
particular those associated with  $gl_n$.
For any root system \cite{OPa, OPc} the Calogero-Moser models are the
natural Hamiltonian systems
\beq
H=\frac{1}{2}\sum_{i} p_i\sp2 +\sum_{\alpha\in\Phi}\, U( \alpha \cdot x);
\label{cmham}
\eeq
where (up to a constant) the potential $U(z)$ is the Weierstrass $\wp$-function
or a degeneration that will be specified below. For the root systems of the
classical algebras a Lax pair may be associated with the models; in the
exceptional setting the existence of a Lax pair is still an open problem.
In fact we do not have a direct proof of
the complete integrability of the Calogero-Moser model associated
with any of the exceptional simple Lie algebras.
Let us consider Lax pairs of the following form \cite{Ca2}:
\beq
L=p\cdot H+\sum_{\alpha\in\Phi} f^{\alpha }E_{\alpha }
,\quad\quad
M=b\cdot H+\sum_{\alpha\in\Phi} w^{\alpha }E_{\alpha },
\eeq
and where the functions $f^{\alpha }$, $w^{\alpha }$ ($\alpha\in\Phi$) are
such that
\beq
f^{\alpha }=f^{\alpha }\left( \alpha \cdot x\right) 
,\quad\quad
w^{\alpha }=w^{\alpha }\left( \alpha \cdot x\right).
\eeq
Then
\beq
\dot L =\dot p \cdot H+\sum_{\alpha\in\Phi} 
         \alpha\cdot \dot x\,f^{\alpha\,\prime}\, E_{\alpha }
\label{ldot}
\eeq
and
\beq
[L,M]=\sum_{\alpha\in\Phi}\biggl(
      \left( \alpha\cdot p\, w^{\alpha } - \alpha\cdot b f^{\alpha }\right)
      E_{\alpha }- f^{-\alpha }w^{\alpha }\alpha\sp\vee\cdot H \biggr)
      +\sum_{\scriptstyle {\beta,\gamma\in\Phi}\atop
             \scriptstyle{\beta+\gamma=\alpha}  }
        c\sp{\alpha}_{\beta\,\gamma} f\sp{\beta}w\sp{\gamma}E_{\alpha } .
\label{lmcomm}
\eeq
We further assume that $b$ is momentum independent.
Upon utilising $\dot x_i=p_i$ and comparing (\ref{ldot}) and (\ref{lmcomm})
we find that the Lax equation $\dot L=[L,M]$ yields the equations
of motion for (\ref{cmham}) provided the following
consistency conditions (for each $\alpha\in\Phi$) are satisfied:
$$
\displaylines{
a)\quad w^{\alpha }=f^{\alpha\,\prime},\hfill\cr
b)\quad
\dot p=-\sum_{\alpha\in\Phi} f^{-\alpha }w^{\alpha}\alpha\sp\vee
=-\sum_{\alpha\in\Phi} f^{-\alpha }f^{\alpha\,\prime}\alpha\sp\vee
=-\frac{d}{dx} \frac{1}{2}\sum_{\alpha\in\Phi}
  {\textstyle\frac{2}{\alpha\cdot\alpha}}f^{-\alpha }
f^{\alpha }\hfill\cr
\quad\phantom{a)\dot p}=-\frac{d}{dx}\sum_{\alpha\in\Phi}
\, U( \alpha \cdot x). \hfill\cr
c)\quad\alpha\cdot b=\sum_{\scriptstyle {\beta,\gamma\in\Phi}\atop
             \scriptstyle{\beta+\gamma=\alpha}  }
        c\sp{\alpha}_{\beta\,\gamma}\frac{f\sp{\beta}w^{\gamma}}
        {f^{\alpha }}
=\sum_{\scriptstyle {\beta,\gamma\in\Phi}\atop
             \scriptstyle{\beta+\gamma=\alpha}  }
        c\sp{\alpha}_{\beta\,\gamma}\frac{f\sp{\beta}f^{\gamma\,\prime}}
        {f^{\alpha }},\hfill
}
$$
The second of the equations determines the potential in terms of the
unknown functions $f\sp\alpha$.
It is the final constraint that is the most difficult to
satisfy. 

Let us now focus on the Lie algebra $gl_n$.
Here $\Phi=\{e_i-e_j,\ 1\leq i\not= j\leq n\}$, where the $e_i$  form
an orthonormal basis of ${\mathbb R}\sp n$.
If $e_{rs}$ denotes the elementary matrix with $(r,s)$-th entry one and
zero elsewhere, then the $n\times n$ matrix representation $H_i=e_{ii}$
and $E_\alpha=e_{ij}$ when $\alpha=e_i-e_j$ gives the usual
representation of $L$. Working with the simple algebra $a_n$ corresponds
to the center of mass frame.
Here the Calogero-Moser models are built from the functions
\beq
f^\alpha =\lambda 
{\sigma(u-\alpha\cdot x)\over{\sigma(u)\sigma(\alpha\cdot x)}}
{e\sp{\zeta(u)\alpha\cdot x}}.
\label{eq:w}
\eeq
These functions satisfy the addition formula
\beq
f\sp\alpha f\sp{\beta\,\prime} -f\sp\beta f\sp{\alpha\,\prime}=
( z_\alpha-z_\beta)f\sp{\alpha+\beta},
\label{eq:addition}
\eeq
where
\beq
 z_\alpha={f\sp{\alpha\,\prime\prime}\over {2 f\sp\alpha}}=
\lambda\wp(\alpha\cdot x) +{\lambda\over2}\wp(u) .
\eeq
Here $\sigma(x)$ and $\zeta(x)=\sigma\sp\prime(x)/ \sigma(x)$
are the Weierstrass sigma and zeta functions \cite{WW}.
The quantity $u$ in (\ref{eq:w}) is known as the spectral parameter.
We find
\beq
U(\alpha\cdot x)=-\frac{\lambda\sp2}{2}\biggl( \wp(\alpha\cdot x)-\wp(u)\biggr)
\eeq
and that
$$b=\frac{1}{2(n+1)}
\sum_{\scriptstyle {\beta,\gamma\in\Phi}\atop
             \scriptstyle{\beta+\gamma=\alpha}  }
        c\sp{\alpha}_{\beta\,\gamma}
z_\beta\, \alpha,
$$
which in components becomes
\beq
b_i=\lambda\sum_{j\ne i} \wp(x_i-x_j).
\eeq
The Weierstrass $\wp$-function includes as degenerations the potentials
\beq
U(z)=\frac{\lambda\sp2}{z\sp2},\quad
\frac{\lambda\sp2}{\sin z\sp2},\quad
\frac{\lambda\sp2}{\sinh z\sp2}.
\eeq
The first of these is the original (rational) Calogero-Moser model
while the second is the Sutherland model \cite{Su2}.

\section{The $gl_n$ Calogero-Moser R-Matrix }
We shall now apply the construction of section two to the
$gl_n$ Calogero-Moser models. Upon examination of (\ref{consgen}) and
(\ref{rmat}) we see that the relevant quantities to
calculate are the matrices $B\sp{\mu\nu}=\{ L\sp\mu, L\sp\nu \}$ and
the components $\Lambda$ and $u$ of $ad(L)$. Using
$\{p_j,f\sp\alpha\}=
 \{p_j,\alpha\cdot q\}f\sp{\alpha\, \prime}=\alpha_j f\sp{\alpha\, \prime}$
we find that
$$
\{L\x L\}=\sum_{\mu,\nu} \{ L\sp\mu, L\sp\nu \} X_\mu\otimes X_\nu
=\sum_{j,\alpha}\alpha_j f\sp{\alpha\, \prime} (H_j\otimes E_\alpha-
E_\alpha\otimes H_j) .
$$
This means that we have 
$$
B=\left(
\begin{array}{cc}
0 & \beta _{i}f^{\beta \,\prime} \\
-\alpha _{j}f^{\alpha \,\prime} & 0
\end{array}
\right)= -B ^{T}
$$
and upon comparison with (\ref{bdef}) we see that $\zeta=\phi=0$.
For the case at hand
\beq
u_{\alpha\, k}=- f\sp{-\alpha}\,\alpha_k ,
\quad\quad
\Lambda _{\beta }^{\alpha }=
\alpha \cdot p\, \delta_{\beta }^{\alpha }+
c_{\alpha -\beta \beta }^{\alpha }\, f^{\alpha -\beta },
\eeq
and it will be convenient to introduce the ($|\Phi|\times\rank\mathfrak{g}$)
matrix
\beq
w_{\alpha\, k}=- f\sp{\alpha\prime}\,\alpha_k .
\eeq
Thus
$
B=\left(
\begin{array}{cc}
0 &- w\sp{T} \\
w & 0
\end{array}
\right)$.
Being quite explicit, if $\alpha=e_i-e_j$ and $\beta=e_r-e_s$ then
\beq
\begin{array}{rl}
 \Lambda:&
   \Lambda\sp{(ij)}_{\ (rs)}  = (p_i-p_j)\delta\sp{i}_{r}\delta\sp{j}_s+
   f(x_i-x_r) \delta\sp{j}_s - f(x_s-x_j)  \delta\sp{i}_r , \\
 u:& u_{(ij), k} = -(\delta_{ik} - \delta_{jk})\, f(x_j -x_i) , \\
 w:& w_{(ij), k} = -(\delta_{ik} - \delta_{jk})\, 
        f\sp\prime(x_i -x_j) ,
\end{array}
\label{explicit}
\eeq
where we adopt the obvious notational shorthand of replacing the
matrix indices for $\alpha=e_i-e_j$ by $(ij)$ and so on.

With these quantities at hand the necessary and sufficient condition
$(C2)$ given by (\ref{consgen}) takes the form
\beq
(C2)\quad\quad
0=u\sp{T}\Lambda ^{-1}w -w\sp{T}\Lambda\sp{-1\, T} u,
\quad\quad\quad\quad\quad\quad
\quad\quad\quad\quad\quad\quad\phantom{xxxx\,}
\label{newc2nocom}
\eeq
which in components becomes
\beq
(C2)\quad\quad
0=
\sum _{\alpha ,\beta }\left(
\alpha_i f^{-\alpha }\left( \Lambda ^{-1}\right) _{\beta }^{\alpha
}f^{\beta\,\prime }\beta _{j}-\alpha _{i}f^{\alpha\,\prime}\left( \Lambda
^{-1}\right) \sp{\beta }_{\alpha }f^{-\beta }\beta _{j}\right).
\phantom{xxxx\,}
\label{newc2}
\eeq
When this is satisfied we have from (\ref{rmat}) that the general R-matrix
is given by
\beq
R=\left( 
\begin{array}{cc}
0 & 0 \\ 
\left( \Lambda ^{-1}\right) _{\beta }^{\alpha }f^{\beta\, \prime }\beta _{j}
& 0
\end{array}
\right) +\left( 
\begin{array}{cc}
p & q \\ 
-\Lambda ^{-1}v p-Fu & -\Lambda ^{-1}v q-F\Lambda ^{T}
\end{array}
\right). 
\label{cmrmat}
\eeq
The second term, which characterises the possible ambiguity in the R-matrix,
was described in  section two.

It is instructive to  consider how the minimal solution
given by the first term of (\ref{cmrmat}) satisfies (\ref{rmatrix}). We have
\beq
R^{\alpha j}=\left( \Lambda ^{-1}\right) _{\beta }^{\alpha }
f^{\beta\,\prime }\beta _{j}
,\quad\quad
R^{ij}=0
,\quad\quad
R^{i\alpha }=0
,\quad\quad
R^{\alpha \beta }=0.
\label{rmatsol}
\eeq
This is to be compared with the previously known R-matrix \cite{AT}
$$
R^{\alpha j}=0
,\quad\quad
R^{ij}=0
,\quad\quad
R^{i\alpha }=-\frac{|\alpha_i|}{2}{f\sp\alpha }
,\quad\quad
R^{\alpha \beta }=\delta_{\alpha+\beta, 0}
 { {f\sp{\alpha \,\prime }}\over f\sp{\alpha }}
$$
which exists only \cite{BS} for the potentials
$U(z)={\lambda\sp2}/{z\sp2}$, ${\lambda\sp2}/{\sin z\sp2}$,
${\lambda\sp2}/{\sinh z\sp2}$.
Examination of the general R-matrix equation (\ref{rmatrix}) yields three 
different equations depending on the range of indices $\{\mu,\nu\}$. For 
$(\mu,\nu)= (i,j),(i,\alpha)$ and $(\alpha,\beta)$ respectively, these are
\begin{eqnarray}
\label{eq:ij}
0&=&\sum_{\alpha}(R\sp{\alpha j}\alpha_i-R\sp{\alpha i}\alpha_j)f\sp{-\alpha},
\\
\label{eq:ia}
\alpha_i f\sp{\alpha\,\prime} &=&
\alpha\cdot p\ R\sp{\alpha i} -\sum_j\alpha_j R\sp{ji}f\sp\alpha +
      \sum_{\beta}(\beta_i f\sp\beta R\sp{-\beta\alpha } +
       f\sp{\alpha -\beta} R\sp{\beta i}c_{\alpha -\beta\,\beta}\sp\alpha) ,
\nonumber  \\
\noalign{\hbox{and}}\cr
\label{eq:ab}
0&=& \sum_{i}(\alpha_i R\sp{i\beta}f\sp\alpha -\beta_i R\sp{i\alpha}f\sp\beta )
  - ( \alpha\cdot p\ R\sp{\alpha\beta}-\beta\cdot p\ R\sp{\beta\alpha})
\nonumber\\
&&  +\sum_{\gamma}(
   R\sp{\gamma\beta} c_{\gamma\, \alpha-\gamma}\sp\alpha f\sp{\alpha-\gamma}
  -R\sp{\gamma\alpha} c_{\gamma\, \beta-\gamma}\sp\beta f\sp{\beta-\gamma}
).\nonumber
\end{eqnarray}
Employing (\ref{rmatsol}) we see that the final two equations are automatically
satisfied. The first equation (\ref{eq:ij}) is less obvious until we realise 
that it just expresses the remaining constraint $(C2)$ necessary for 
a solution to exist. This identification follows upon using 
$R^{\alpha j}=\left( \Lambda ^{-1}\right) _{\beta }^{\alpha }
f^{\beta\,\prime }\beta _{j}$.

At this stage we have reduced the existence of an R-matrix for
Calogero-Moser systems to that of a constraint equation:
\begin{thm}
The elliptic Calogero-Moser system has R-matrix (\ref{rmatsol}) if and only
if (\ref{newc2}) --or equivalently (\ref{newc2nocom})-- is satisfied.
\end{thm}

We remark that (\ref{newc2nocom}) is again of the form (\ref{matrixeqn})
for the (nonsquare) matrix
$\tilde A=\Lambda ^{-1\,T} u$ and $\tilde B=0$, where now we wish to show that
$\tilde X=w$  is a solution.  The general theory applies and
as $\tilde B=0$ the constraints are automatically satisfied.
One discovers in this situation
that the requirement  $P_2\sp{T}ZP_2$ be symmetric 
is equivalent to the symmetry of the matrix 
$w\sp{T} \Lambda ^{-1\,T} u$.

\section{The Constraint}

It remains to analyse the constraint equation (\ref{newc2}).
Although the
inverse matrices here look somewhat daunting we may use the cofactor
expansion of an inverse to give
$$
\left\vert\begin{array}{cc} 0&k\\ l &\Lambda \end{array}\right\vert
=-|\Lambda |\ k\sp{T}\Lambda ^{-1}l.
$$
Thus (\ref{newc2}) is equivalent to showing that (for each $i,j$) the
$(|\Phi|+1)\times(|\Phi|+1)$ determinants satisfy
\beq
\left\vert\begin{array}{cc}
0&\alpha _{i}f^{-\alpha}\\
\beta_{j}f^{\beta\,\prime}&\Lambda
\end{array}\right\vert
=
\left\vert\begin{array}{cc}
0&\alpha _{j}f^{-\alpha}\\
\beta_{i}f^{\beta\,\prime}&\Lambda
\end{array}\right\vert
,
\label{c2prov}
\eeq
where\footnote{Note the adjugate matrix of $\Lambda$ involves the transpose
of the cofactors and hence the perhaps puzzling interchange of rows and columns
here.}
$\Lambda=\left(\Lambda_{\beta\alpha}\right)$.
To be quite explicit we wish to show that (for each $i,j$)
\beq
\left\vert\begin{array}{cc}
0&u_{(rs),i}\\
w_{(kl),j}&\Lambda\sp{(kl)}_{\ (rs)}\\
\end{array}\right\vert
=
\left\vert\begin{array}{cc}
0&u_{(rs),j}\\
w_{(kl),i}&\Lambda\sp{(kl)}_{\ (rs)}\\
\end{array}\right\vert
,
\label{c2explicitprov}
\eeq
where  $\Lambda$ (to be invertible), $u$ and $w$ are given 
by (\ref{explicit}), the
indices $(kl)$, $(rs)$ run over ordered distinct pairs and the
functions being considered are given by 
$ f(x) =
{\sigma(u- x)\over{\sigma(u)\sigma( x)}}
{e\sp{\zeta(u) x}}.
$
Actually, because of the symmetry of the problem, it suffices to show that
(\ref{c2prov}) holds for any two indices $i\ne j$ (it clearly holding for $i=j$)
and we may take these for example to be $i=1$, $j=2$.

We are unable to prove  (\ref{c2prov}) in generality. Symbolic
manipulation has verified it true for small numbers of particles and it
has satisfied extensive numerical checks.
At present we can only present it as a conjecture.
The conjectured identity appears new.

We remark that in the present setting one can show that
for arbitrary functions $f\sp{\alpha}$ (for which $\Lambda$ is invertible)
\beq
0=
\left( u^{T}\Lambda ^{-1}v \right) _{ij}=\sum_{\alpha,\beta }\alpha\sp\vee
_{i}L^{-\alpha }\left( \Lambda ^{-1}\right) _{\beta }^{\alpha
}L^{\beta }\beta _{j}
=
\left\vert\begin{array}{cc}
0&\alpha _{i}f^{-\alpha}\\
\beta_{j}f^{\beta\,}&\Lambda
\end{array}\right\vert .
\label{rankprov}
\eeq
Whereas (\ref{rankprov}) is true for any functions $f\sp{\alpha}$
equation (\ref{c2prov}) will only hold for a more restricted class 
of functions. The constraint requires that 
functions of the form (\ref{eq:w}) satisfy (\ref{c2prov}).

\section{Discussion}
This paper has been devoted to the  construction of a spectral parameter
independent R-matrix for the elliptic Calogero-Moser models.
Previous work has shown that no momentum independent and spectral parameter
independent R-matrix exists for the models for more than three particles. 
By viewing the R-matrix
equation as a particular case of the general matrix equation
(\ref{matrixeqn}) we are able to give
necessary and sufficient conditions for a (generally momentum
dependent) R-matrix to exist. No recourse to special ans\"atze is needed
and the general form of the R-matrix can be specified.
The elliptic Calogero-Moser model 
has R-matrix (\ref{rmatsol}) if and only
if (\ref{newc2}) --or equivalently (\ref{newc2nocom})-- is satisfied.
The desired R-matrices existence has thus been reduced to  the validity of
a single constraint. This constraint may equally be cast as the
equality betwen two determinants (\ref{c2prov}) involving elliptic
functions. (We have unpacked most of the Lie algebra notation in the
explicit form (\ref{c2explicitprov}).)
Such an identity  appears new. Unfortunately we have
been unable to prove (\ref{c2prov}) in generality and we  present it here
as a conjecture. 

\section{Acknowledgements}
This material was presented at the CRM \lq Workshop on Calogero-Moser-
Sutherland Models\rq\ (Montreal, March 1997) and ICMP'97 (Brisbane, July 1997).
I have benefited from comments by
J. Avan, J. Harnad, A.N.W. Hone,
I. Krichever, V. Kuznetsov, M. Olshanetsky and E. Sklyanin.


\begin{thebibliography}{99}

\bibitem{Arnold} V.I. Arnold, Mathematical Methods of Classical Mechanics,
        Springer-Verlag, New York 1978.

\bibitem{AT} J. Avan and M. Talon,
        {\it Classical R-matrix structure for the Calogero model},
        Phys. Lett. {\bf B303}, 33-37 (1993).

\bibitem{BV} 0. Babelon and C.M. Viallet,
        {\it Hamiltonian Structures and Lax Equations},
        Phys. Lett. {\bf B237}, 411-416 (1990).


\bibitem{AG}Adi Ben-Israel and Thomas N.E. Greville,
        { Generalized inverses : theory and applications},
        Krieger, Huntington N.Y. 1974.  

\bibitem{BS}H.W. Braden and Takashi Suzuki,
        {\it $R$-matrices for Elliptic Calogero-Moser Models},
        Lett. Math. Phys. {\bf 30}, 147-158 (1994).

\bibitem{Br1}H.W. Braden,
        {\it The Equation $A\sp{T}X-X\sp{T}A=B$},
        Edinburgh Preprint MS-97-007.

\bibitem{Br2}H.W. Braden,
        {\it R-matrices and Generalized Inverses},
        J. Phys. {\bf A 30}, L485-L493 (1997) {\tt solv-int/9706001}.

\bibitem{Ca2}F. Calogero, {\it On a functional equation connected with
      integrable many-body problems}, Lett. Nuovo Cimento 16, 77-80 (1976).

\bibitem{Ca}S.R.~ Caradus,
        {\it Generalized inverses and operator theory},
        Queen's papers in pure and applied mathematics No. 50,
        Queen's University, Kingston Ont. 1978.  

\bibitem{flaschka}H. Flaschka,
        {\it A Remark on Integrable Hamiltonian Systems},
         Phys Lett {\bf A 131}, 505-508 (1988).

\bibitem{KNS} V.B. Kuznetsov., F.W. Nijhoff and E.K. Sklyanin,
        {\it Separation of variables for the Ruijsenaars system}, 
        March 1997, accepted in {\it Commun. Math. Phys.};
        {\tt solv-int/9701004}.

\bibitem{Liouv}J. Liouville,
       {\it Note sur les \'equations de la dynamique},
       J. Math. Pures Appl. {\bf 14} 137-138 (1855).

\bibitem{OPa}M.A. Olshanetsky and A.M. Perelomov,
     {\it Classical Integrable finite-dimensional systems related to
      Lie Algebras}, Phys. Reps. 71, 313-400 (1981).

\bibitem{OPc}M.A. Olshanetsky and A.M. Perelomov,
     {\it Completely integrable Hamiltonian systems connected with
       semisimple Lie algebras},
       Invent. Math. 37, 93-108  (1976).

\bibitem{Pr}R.M.~Pringle and A.A. Rayner,
      { Generalized inverse matrices with applications to statistics},
      Griffins statistical monographs and courses No. 28,
      Charles Griffin, London 1971.

\bibitem{RM}C. Radhakrishna Rao and Sujit Kumar Mitra,
      { Generalized Inverse of Matrices and its Applications}
      John Wiley and Sons, New York 1971.

\bibitem{STS} M.A. Semenov-Tian-Shansky,
      {\it What is a classical r-matrix?},
      Funct. Anal. Appl. {\bf 17} 17-33 (1983).

\bibitem{Sk1} E.K. Sklyanin,
       {\it Dynamical r-matrices for the elliptic Calogero-Moser Model},
       St Petersburg Math. J. {6}, 397-406 (1995).

\bibitem{Sk2} E.K. Sklyanin,
       {\it Separation of Variables: New trends},
       Prog. Theor. Phys. Suppl. {\bf 118} 35-60 (1995).

\bibitem{Su2}B. Sutherland, {\it Exact results for a quantum many-body
      problem in one dimension}, Phys. Rev. A4, 2019-2021 (1971);
      II Phys. Rev. A5, 1372-1376 (1972).

\bibitem{WW}E.T. Whittaker and G.N. Watson, { A Course of Modern Analysis},
     Cambridge University Press 1927.

\end{thebibliography}
\end{document}